\documentclass[]{spie}  

\newcommand{\um}{$\mu$m}
 
\usepackage{amsmath,amsfonts,amssymb}
\usepackage{graphicx}

\usepackage[colorlinks=true, allcolors=blue]{hyperref}

\title{Enhancing the performance and capabilities of the MIRI instrument on JWST}

\author[a]{S. Kendrew}
\author[b]{M. Engesser}
\author[c]{K. Rowlands}
\author[b]{A. Noriega-Crespo}
\author[a]{M. Decleir}
\author[b]{H. Diamond-Lowe}
\author[b]{M. Regan}
\author[b]{J. Aguilar}
\author[c]{S. Alberts}
\author[c]{T. Bell}
\author[b]{M. Cracraft}
\author[b]{A. Dyrek}
\author[b]{K. Gordon}
\author[b]{D. Hines}
\author[b]{B. J. Holler}
\author[c]{K. Larson}
\author[b]{D. Law}
\author[b]{K. Murray}
\author[b]{B. Nickson}
\author[b]{B. O'Sullivan}
\author[b]{A. Petric}
\author[b, d, e]{B. Sargent}
\author[b]{S. Shenoy}
\author[b]{G. C. Sloan}
\author[b]{B. Trahin}
\author[b]{I. Wong}

\affil[a]{European Space Agency, Space Telescope Science Institute, Baltimore MD, USA}
\affil[b]{Space Telescope Science Institute, Baltimore MD, USA}
\affil[c]{AURA for ESA, Space Telescope Science Institute, Baltimore MD, USA}
\affil[d]{SETI Institute, USA}
\affil[e]{Johns Hopkins University, Baltimore MD, USA}

\authorinfo{Further author information: \\ E-mail: skendrew@stsci.edu}

\begin{document} 
\maketitle

\begin{abstract}
MIRI, the Mid-Infrared Instrument on the James Webb Space Telescope, is the only instrument sensitive to wavelengths longward of 5~\um\ on the
observatory. In this regime, MIRI brings order-of-magnitude improvements over past instruments and missions in a wavelength range that is challenging
or impossible to access from the ground. As such, MIRI occupies a unique parameter space, offering unparalleled capabilities in all areas of astrophysics,
from the Solar System to the most distant galaxies in the Universe. We continue to make operational improvements to MIRI, four years into its operational
lifetime, to create new observing opportunities, improve performance, and enhance its scientific return. In this paper we will describe several such
improvements, and their anticipated impact on MIRI and JWST science.
\end{abstract}


\section{INTRODUCTION}
\label{sec:intro}  

MIRI~\cite{2015PASP..127..595W, 2023PASP..135d8003W}~is one of four science instruments on the James Webb Space Telescope (JWST). Sensitive to wavelengths from $\sim$5~to 28.5~$\mu$m, it provides the following mid-infrared capabilities for JWST:

\begin{itemize}
\item Imaging~\cite{2015PASP..127..633B, 2024A&A...689A...5D}
\item Low-Resolution Spectroscopy (LRS; slit and slitless)~\cite{2015PASP..127..623K}
\item Coronagraphic imaging~\cite{2015PASP..127..633B, 2022A&A...667A.165B}
\item Wide-Field Slitless Spectroscopy (new for Cycle 5)
\item Medium-Resolution Integral Field Spectroscopy (MRS)~\cite{2015PASP..127..646W, 2023A&A...675A.111A}
\end{itemize}

The imaging, coronagraphic imaging, low-resolution and wide-field slitless spectroscopy modes make up the MIRI Imager, sharing a single 1032 $\times$ 1024 Si:As Impurity Band Conduction (IBC) detector~\cite{2015PASP..127..584R, 2015PASP..127..675R}. The imager filter wheel contains the imaging science filters, coronagraphic filters, and a ZnS/Ge double prism assembly (P750L) providing R $\sim$ 40 -- 160 dispersion from 5 to 14~\um; in addition to engineering optics. The MRS mode uses two 1032 $\times$ 1024 Si:As IBC detectors, with R $\sim$ 1500 -- 3500, 4.9 -- 27.9~\um\ integral field spectroscopy performed by an image slicer combined with a set of dichroics and gratings. The full wavelength range is split into 4 channels and 3 bands; each exposure provides coverage of all 4 channels in one of 3 bands. 

MIRI detectors are read out using the MULTIACCUM readout scheme, which was adopted for all JWST instruments. The array is read out non-destructively at intervals defined by three key parameters during each exposure\cite{2015PASP..127..675R, 2023PASP..135g5004M}: (i) \textit{nsample}, the number of samples taken per pixel (this parameter is implicit in the definition of a particular readout pattern, not chosen by the user); (ii) \textit{ngroup}, the number of groups during an integration, where a group is the product of cycling through all the pixels; and (iii) \textit{nints}, the number of integrations during an exposure, where an integration is defined as the time between resets. Specific to MIRI is the inclusion of an additional reset between integrations, which was found to reduce reset effects in multi-integration exposures\cite{2023PASP..135g5004M}.

The MIRI detector readout follows the MULTIACCUM readout scheme, where pixels are read out non-destructively until the array is reset. Each read step constitutes a frame or group, and each sequence of groups bracketed by resets is an integration. Note that for the majority of MIRI's observations, groups always contain a single frame, so in most cases these terms are equivalent; this is not the case for JWST's other instruments. The calibration pipeline uses each up-the-ramp read (in DN) to fit a slope to the accumulating charge (in DN/s). An exposure groups together one or more integrations, and is written out to a file. 

Four years into JWST operations we are continuously improving our understanding of the instrument's performance and how to optimize data quality and calibration. In this paper we describe a number of projects in progress at the Space Telescope Science Institute (STScI), the host of the JWST Mission Operations Center, that will bring new capabilities and performance improvements for MIRI. These projects constitute major efforts for the MIRI Instrument Team at STScI, in collaboration with the engineering and software teams, and complement the ``routine'' support for MIRI - user support, technical reviews, calibration and documentation. Two of these projects are the topic of dedicated papers in this conference; these will only be briefly summarised here. 

\section{Wide-Field Slitless Spectroscopy Mode}\label{sec:wfss}

The most significant enhancement to MIRI’s capabilities is the introduction of a new observing mode to perform low-resolution Wide-Field Slitless Spectroscopy (WFSS). This combines the R $\sim$ 100, 5 -- 14 \um\ dispersion by the double prism (already used for the LRS modes) with the large Imager field of view. When the double prism is selected in the filter wheel and the full array is read out, as is typical for LRS fixed-slit observations, any sources located in the Imager portion of the field are also dispersed. As a result, slitless spectra are serendipitously recorded in many or most LRS fixed-slit observations. Enabling this as a dedicated mode was therefore a natural extension of MIRI's capabilities. Petric et al provide a comprehensive overview of this mode in these proceedings. 

The WFSS mode is available from Cycle 5, which started in July 2026, with implementation for parallel modes from Cycle 6 (starting July 2027).

\section{New subarrays to mitigate EMI noise}\label{sec:new_subarrays}

Early in the mission, a spatially and temporally varying periodic noise signal was noted to affect a number of subarrays on the Imager detector~\cite{2023PASP..135c8002B, 2023PASP..135g5004M}. The cause of this was identified as Electromagnetic Interference (EMI) at certain frequencies, when these are out of phase with the interval between pixel reads. The most prominent frequency is at 390 Hz, and affects the SLITLESSPRISM, SUB128, SUB64 and all MASK* coronagraphic subarrays. The effect is particularly prominent in time-series observations (TSOs)\cite{2023PASP..135c8002B}, which typically contain hundreds or even thousands of integrations, processed individually to construct the time series---making the moving noise pattern highly visible. The high stability requirement for TSOs makes such a varying noise pattern highly undesirable. A detailed description of this EMI phenomenon is given in Brandt et al~\cite{2025jwst.rept.9091B}. 

Since Build 10.2 (May 2024) the JWST calibration pipeline~\cite{2024zndo..12556702B} has included the {\tt emicorr} step to fit and remove this pattern; the algorithm is also described in Brandt et al~\cite{2025jwst.rept.9091B}. The pattern can however be prevented from arising by designing subarrays such that the pixel read time will be in phase with the dominant noise frequencies. To this end, the decision was made to create and implement a new set of subarrays to replace those affected by the dominant 390 Hz EMI noise. For SUB128 and SUB64, the new subarrays were designed to be at the same location, with subtly different shape to bring the readout time in phase with the noise signal. The frame read times for SUB128\_{IP} and SUB64\_{IP} (where ``IP'' stands for ``In Phase'') are within 5\% of those of their out-of-phase counterparts, ensuring that observations can be switched between old and new versions with very little time change. Due to the tight tolerances on alignment with the coronagraphic masks, the MASK* subarrays used for coronagraphic imaging were left unchanged. 

For the LRS mode, several other issues were taken into consideration for the design of the new subarrays. The SLITLSSPRISM subarray, the subarray used for most TSOs with MIRI is, in addition to the EMI noise, also affected by a problematic ``shadow region.'' This region constitutes pixels that are located behind the focal plane mask, only receiving light when the P750L double prism is selected in the filter wheel. The rest of the subarray receives some level of illumination continuously regardless of the filter selected in the wheel. At the start of a new exposure, the ``shadow region'' pixels therefore often have a significantly different illumination history than the others in the subarray, leading to a markedly different settling pattern that is difficult to calibrate. This issue was first reported by Bell et al.~\cite{2023arXiv230106350B}, affects around 20\% of LRS TSOs, and has proven challenging to characterize, predict, or prevent. The pixel region corresponds to a wavelength range of $\sim$10.6 -- 11.8~\um, a scientifically valuable region with good throughput for the prism. To mitigate the impact of this phenomenon, the new in-phase version of the subarray, SLITLESSPRISM\_IP was moved on the array to push the problematic pixel region to longer wavelengths ($>$ 12.2~\um), where the instrument throughput is much lower.

In addition, two new subarrays were created for the LRS mode. The first, SLITLESSPRISM\_IPS (where ``S'' designates ``short'') increases the dynamic range of the slitless LRS mode by 24\% at the expense of wavelength coverage; while the LRS double prism disperses out to $\sim$14~\um, albeit with a steep drop-off in transmission past 10~\um, the SLITLESSPRISM\_IPS  subarray cuts off coverage around 12.5~\um. As a result, this smaller, faster subarray is entirely unaffected by the ``shadow region.'' The second new subarray provides a faster readout option for the fixed-slit mode of LRS. The SUBSLIT subarray includes the slit region and a small portion of the Imager field of view to allow for target acquisition exposures, offering a factor 9 increase in dynamic range compared to the FULL array readout for the current LRS slit observations. Both of these new subarrays will be unaffected by 390 Hz EMI noise, and offer substantial new capabilities to the instrument. 

\begin{figure}
\centering
\includegraphics[width=10cm]{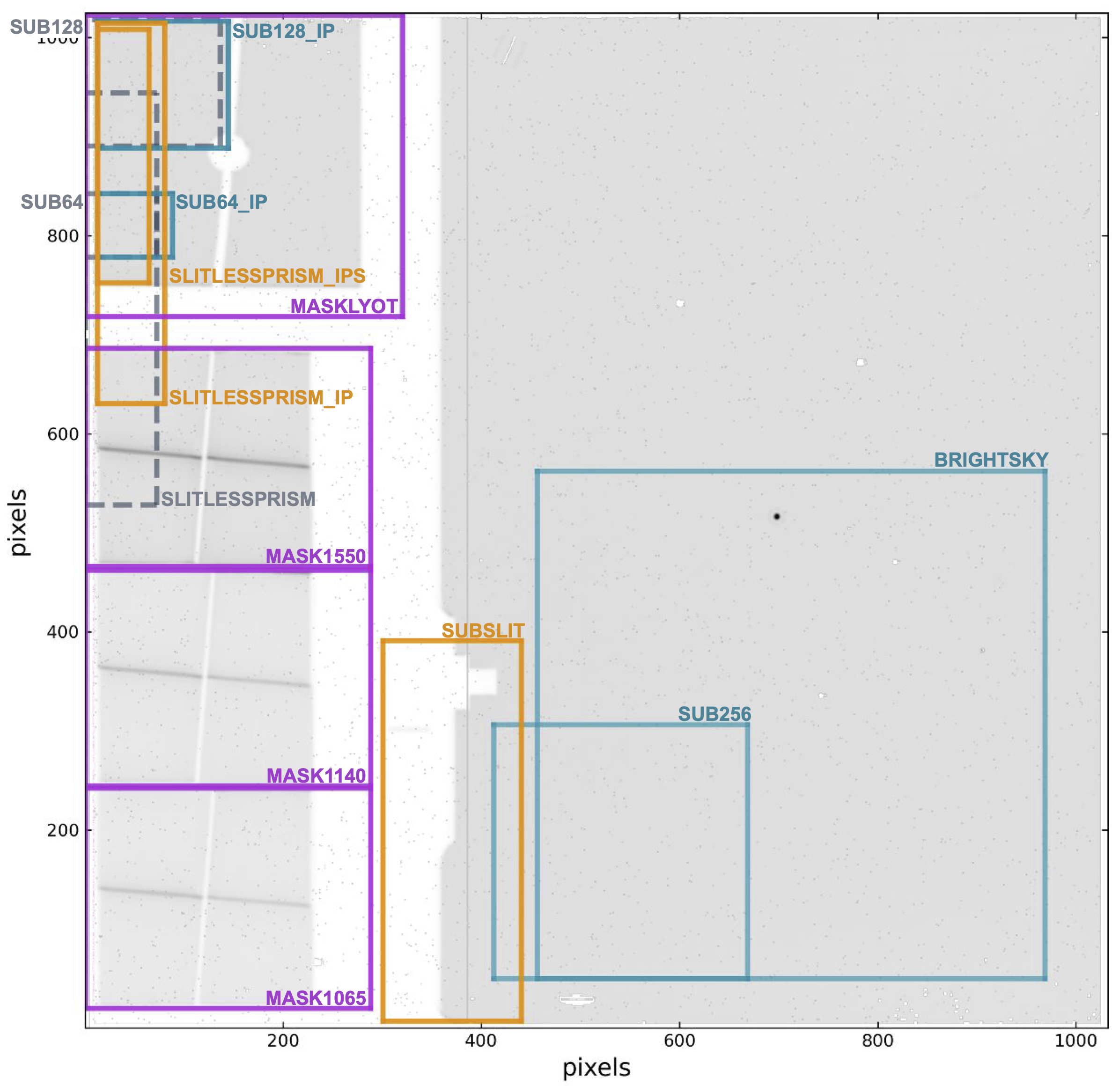}
\caption{Subarray locations on the imager detector for imaging (teal), LRS (gold), and coronagraphy (purple). All illustrated subarrays are available in Cycle 6. Subarrays in dashed grey lines will be retired in future cycles. Figure replicated from the \href{https://jwst-docs.stsci.edu/jwst-mid-infrared-instrument/miri-observing-modes/miri-imaging}{JWST Documentation}. }\label{fig:subarrays}
\end{figure}

The current and new subarrays are shown in Fig.~\ref{fig:subarrays}, and sizes and readout times for all subarrays, new as well as current, are listed in Table~\ref{tab:subarrays}. A transition period is planned for Cycle 6 to provide observers with continuity from previous cycles, before deprecating the old subarrays in Cycle 7. Calibration observations are planned for 2026/2027 to bring these new subarrays to science-readiness. 

EMI noise will still occur at lower levels at other frequencies (see Brandt et al.\cite{2025jwst.rept.9091B}); the \texttt{emicorr} step in the calibration pipeline will continue to correct these in all MIRI data as needed. 

\begin{table}[]
    \centering
    \begin{tabular}{|c|c|c|c|}
    \hline
       Subarray  & Rows $\times$ columns & Read time (s) & Observing mode\\
       \hline
        FULL & 1024 $\times$ 1032 & 2.775 & Imaging/Slit LRS/WFSS/Coronagraphy\\
        BRIGHTSKY & 512 $\times$ 512 & 0.865 & Imaging\\
        SUB256 & 256 $\times$ 256 & 0.300 & Imaging\\
        SUB128 & 128 $\times$ 128 &  0.119 & Imaging\\
        SUB128\_IP(*) & 128 $\times$ 132 & 0.120 & Imaging\\
        SUB64 & 64 $\times$ 72 &  0.085 & Imaging\\
        SUB64\_IP(*) & 64 $\times$ 76 &  0.087 & Imaging\\
        SLITLESSPRISM & 416 $\times$ 72 & 0.159 & Slitless LRS\\
        SLITLESSPRISM\_IP(*) & 384 $\times$ 68 & 0.156 & Slitless LRS\\
        SLITLESSPRISM\_IPS(*) & 256 $\times$ 52 & 0.118 & Slitless LRS\\
        SUBSLIT(*) & 384 $\times$ 140 & 0.279 & Slit LRS\\
        \hline
    \end{tabular}
    \caption{Overview of the MIRI Imager subarrays affected by the updates. The subarrays marked with (*) are new for Cycle 6 and unaffected by 390 Hz EMI noise. Their out-of-phase counterparts are also listed for comparison. Note the BRIGHTSKY and SUB256 subarrays are not affected by 390 Hz noise, so they remain unchanged and available as in previous cycles. }
    \label{tab:subarrays}
\end{table}

\section{FASTGRPAVG8 readout mode for science observations}\label{sec:fastgrpavg}

In Section~\ref{sec:intro} we introduced the MULTIACCUM readout scheme used by MIRI and the other JWST instruments. MIRI's default readout pattern is FASTR1, where \textit{nsample} is 1 and \textit{nreset} is also 1; in FASTR1, each pixel is addressed just once, and one additional reset is executed between integrations. The full-frame readout time in FASTR1 is 2.775 s. 

For very long or low-background observations, the SLOWR1 mode was implemented. In SLOWR1, the pixel is sampled 8 times in a 9-sample wide window\cite{2023PASP..135g5004M}. The first sample is ignored and the remaining 8 samples are averaged to output a single result (and the \textit{nsample} parameter is set to 9). The resulting readout time for a single SLOWR1 group is 23.89 s. The readout noise for SLOWR1 was expected to be reduced by a factor of $\sqrt{8}$ compared with FASTR1, and in addition reduces the data volume for observations where multiple detectors are read out simultaneously, e.g., the MRS mode or parallel observations with another instrument. MIRI is read noise dominated in the shortest-wavelength imager filters (F560W and F770W) and in the MRS short-wave channel.

\begin{figure}
    \centering
    \includegraphics[width=0.5\linewidth]{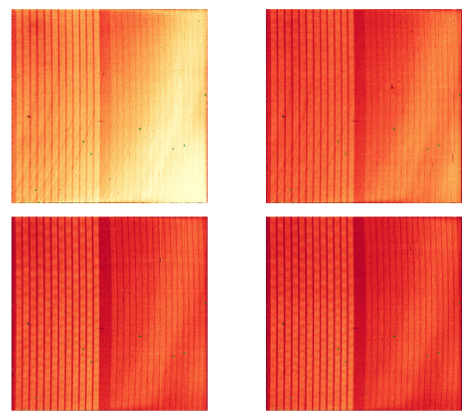}
    \caption{Sequence of the MIRI MRS long-wavelength detector (MIRIFULONG) starting right after a switch into SLOWR1 mode; the scale is 0-2 DN/s. Time goes left to right and top to bottom. The strong arc seen on the lower right of the images is a thermal artifact often seen after the FASTR1 to SLOWR1 switch. Additionally, there is a strong even/odd row pattern that decreases with time. The exposures are separated by 2.5 minutes. Figure from Regan~\cite{regan_slowr1}.}
    \label{fig:slowr1}
\end{figure}

Regan\cite{regan_slowr1} performed a study to characterize the read noise differences between readout patterns. These investigations showed the readout noise for SLOWR1 to be higher than expected; in addition, the switching between FASTR1 and SLOWR1 causes thermal settling artifacts, particularly visible in the MRS detectors. As such, a program using FASTR1 that follows a program using SLOWR1 can be affected by these undesirable artifacts. An example of the time sequence in an MRS array following the switch is shown in Fig.~\ref{fig:slowr1}.

MIRI offers an alternative low data volume readout strategy, with a set of  FASTGRPAVG* readout modes, which average 4, 8, 16, 32 or 64 single-read frames (i.e., FASTRGRPAVG, FASTGRPAVG8, FASTGRPAVG16, etc). Contrary to the SLOWR1 mode, FASTRGRPAVG* readout modes do not require a state change in the detector. Each pixel is read just once, as in the FASTR1 mode, and the frames are combined outside of the instrument, prior to the data being written to the on-board data storage. These modes were implemented specifically for Target Acquisition (TA), for which there is a limit of 10 groups that can be held in memory for the centroiding algorithm. As such they enable longer TA sequences on fainter targets. As of Cycle 4, these modes were only offered for TA. 

\begin{figure}
    \centering
    \includegraphics[width=\linewidth]{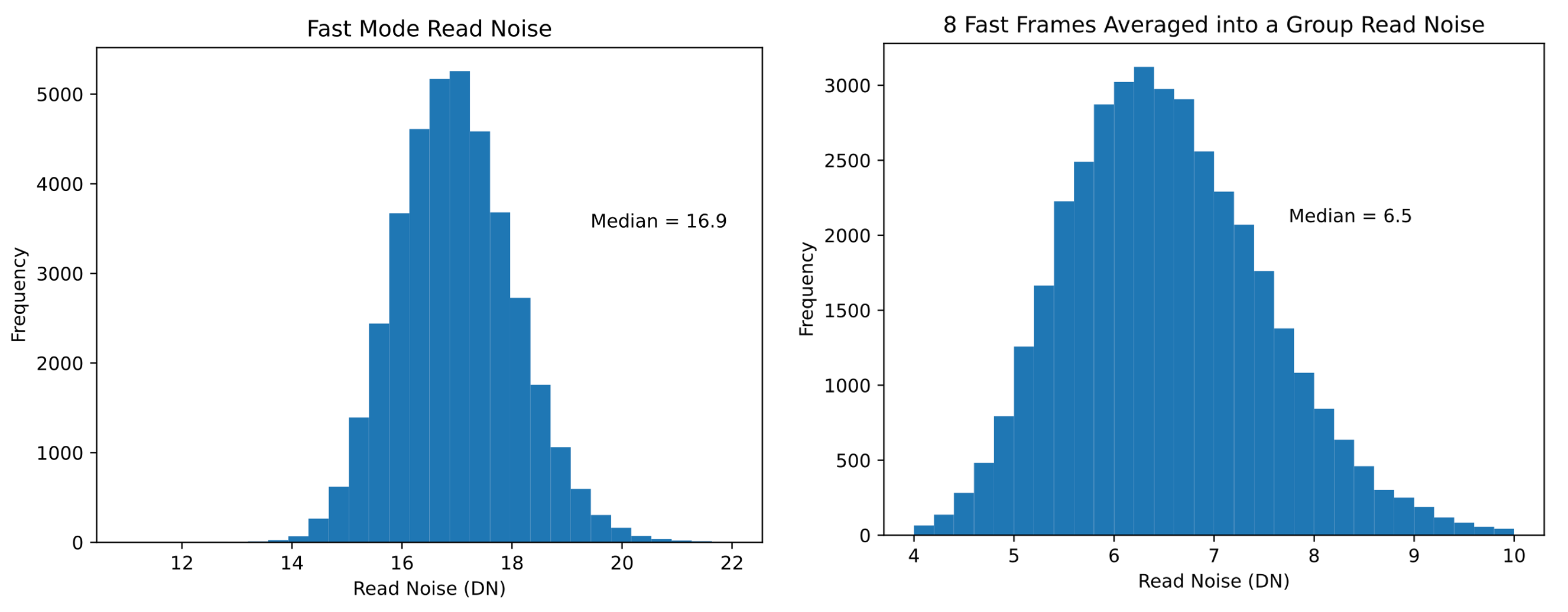}
    \caption{Comparison of read noise measured for the FASTR1 and FASTRGRPAVG8 readout patterns, illustrating the improvement. This will particularly benefit short-wavelength imaging (F560W and F770W) and short-wavelength spectroscopic observations with the MIRI MRS. Figures from Regan\cite{regan_slowr1}.}
    \label{fig:regan_readnoise}
\end{figure}

As the FASTGRPAVG* modes achieve the same reduction in data volume compared with FASTR1 and a $\sqrt{N}$ reduction in readout noise (median of 6.5 DN, see Fig.~\ref{fig:regan_readnoise})\footnote{all read noise numbers quoted here are subject to change based on an ongoing analysis}, while avoiding the thermal settling artifacts related to the FASTR1-SLOWR1 switch, they provide an attractive alternative to the SLOWR1 pattern. The FASTGRPAVG8 mode in particular is a close alternative to the SLOWR1 mode, and the decision was made to begin using this mode for the observations where data volume mitigation is required. This change will take effect formally from JWST Cycle 6 (July 2027); the SLOWR1 mode will be offered for a transition period, before being formally deprecated. All FASTGRPAVG* modes will remain available for Target Acquisition for MIRI LRS, MRS, and Coronagraphic imaging. 

\section{Dithered Target Acquisition Exposures}\label{sec:tadither}

TA is a crucial part of the observation for several of MIRI's modes. For fixed-slit Low Resolution Spectroscopy (LRS), the slit aperture is just 0.5'' across, and well-centered target placement is essential for accurate calibration. For coronagraphic imaging, the tolerance on placement of the target on the vertex position of the 4-quadrant phase masks (4QPMs) is $\sim$5 mas\cite{2022A&A...667A.165B}. Finally, for the Medium-Resolution Spectrometer, high-precision control over target placement of point sources can aid in removal of fringing patterns~\cite{2025A&A...697A..58G}. For JWST instruments, the user specifies a TA exposure configuration, using either the science target itself or a nearby bright point source. A generic on-board algorithm~\cite{valenti2024} performs a center-of-mass centroiding to determine the source location. The small-angle maneuver required to offset this position to the pointing position in the science aperture is computed and executed. This TA procedure is routinely executed for MIRI with positive outcomes in the majority of observations. 

An investigation by Decleir et al (2026, in prep) into the successes and failures of TA sequences for the fixed-slit LRS mode determined a success rate of 88\%, based on 431 observations executed between the start of operations (July 2022) and late May 2026. They find a majority of failures are due to user error; the most common types of user error are:

\begin{itemize}
    \item reaching insufficient SNR in the TA exposure due to computation errors, uncertain target brightness in the chosen filter;
    \item saturating on the TA target in the chosen filter and exposure duration;
    \item using a TA target which has a brighter nearby companion or bright extended emission, causing the algorithm to centroid on the companion or extended emission instead;
    \item coordinate or proper motion errors, causing the target to fall outside of the TA region.
\end{itemize}

Failures or poor data quality due to user error are not usually granted an opportunity to repeat the observation. 

\begin{figure}
    \centering
    \includegraphics[width=\linewidth]{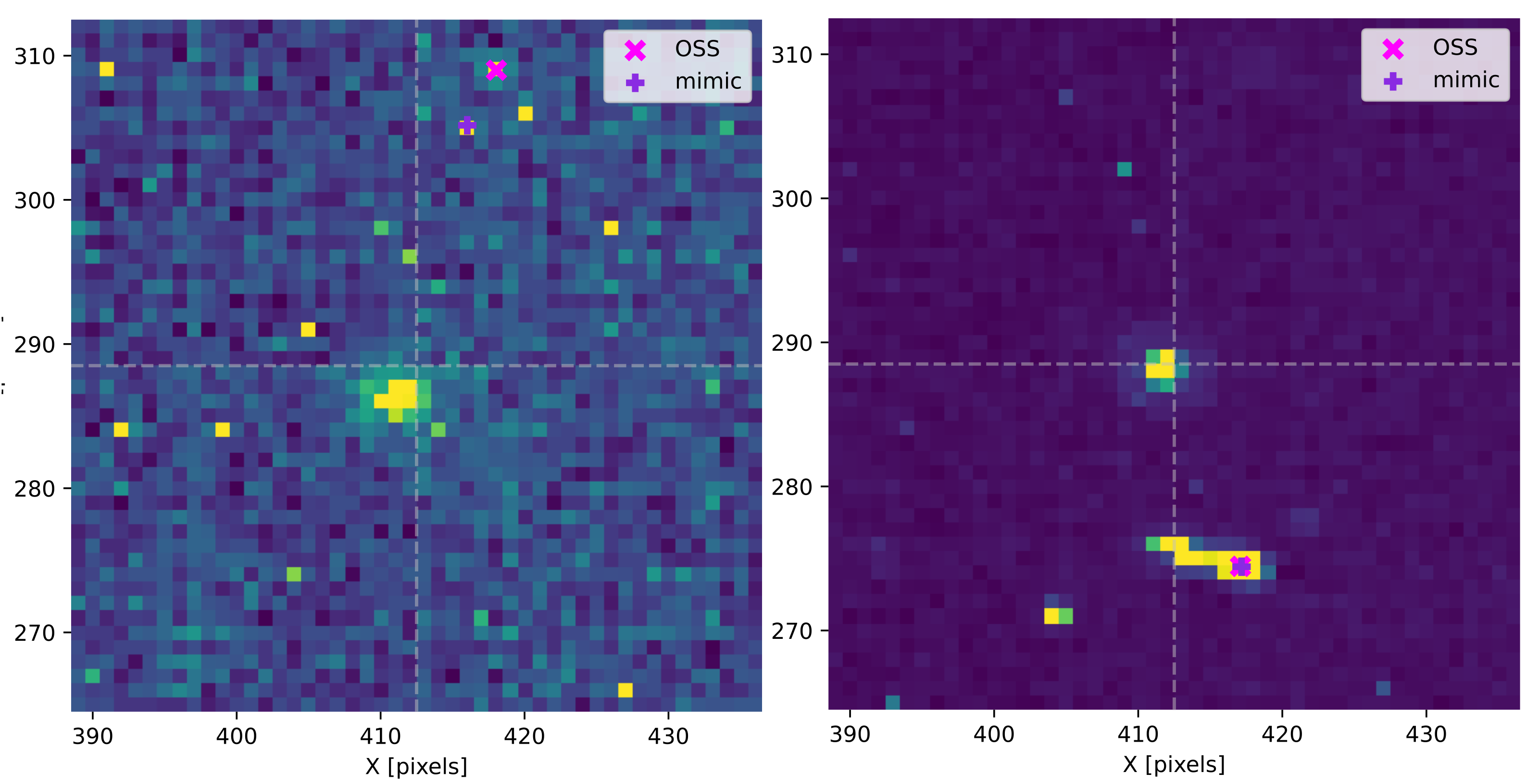}
    \caption{Example of TA sequences that failed due to an unmasked hot pixel (left) and a cosmic ray (right). The magenta $\times$ labelled ``OSS'' marks the location where the on-board algorithm identified the centroid; the purple + shows the centroid as identified by a local analog script used to study the failures. Figure adapted from Decleir et al (2026, in prep).}
    \label{fig:failed_tas}
\end{figure}

A small number of additional TA failures are caused by unmasked hot pixels or cosmic ray hits. We show examples of such cases in Fig.~\ref{fig:failed_tas}. The figures show the TA region of interest (ROI), the 48 $\times$ 48 px region into which the target is placed for centroiding. The dashed lines indicate the nominal centre of the ROI. Both panels are taken from an LRS slit TA sequence, but all MIRI modes using TA have dedicated TA ROIs on the Imager array. In these cases users are typically granted an opportunity to repeat the observation. While these are rare (just $\sim$5\% of observations based on the studied sample), they can nonetheless result in failed observations, poor data quality, gaps in the schedule, and general time inefficiency; and we strive to avoid these occurrences. 

Bad pixels are mitigated in TA with the aid of bad pixel masks uploaded to the spacecraft. The algorithm uses this mask to replace the bad pixel values by a median of the surrounding pixels. However, analysis has shown that the MIRI imager detector sees an increase in anomalous ``rogue'' pixels of (on average) $\leq$ 9 pixels/day~\cite{engesser2025}, and their locations are unpredictable. This puts a significant burden on active monitoring and regularly re-generating and re-delivering the on-board mask.

Cosmic rays (CRs) are entirely unpredictable, and difficult to mitigate. The on-board TA algorithm performs a very simple conversion from raw frames in DN to a slope image in DN/s via a group differencing method, where the TA image is constructed from the minimum value of (middle group $-$ first group)  and (last group $-$ middle group)\footnote{for MIRI specifically the last group is disregarded due to the last frame pulldown effect\cite{2023PASP..135g5004M}; the second group difference uses the last-but-one frame instead}, calculated pixel by pixel. If the jump in counts from a CR falls entirely in the first or second half of the ramp, the slope image construction method will eliminate it from the final TA image. If however the CR hits near the middle of the ramp and the count increase associated with the CR's energy straddles the mid-point of the ramp, the resulting artifact will persist in the TA image and the algorithm risks placing the centroid on the CR hit. An example of this is shown in the right-hand panel of Fig.~\ref{fig:failed_tas}. 

Both the bad pixel and CR risks to successful TA can be greatly mitigated by performing a dither pattern for the TA exposure, rather than take a single image. By taking not one, but three exposures, shifted by a small number of pixels, CR artifacts and unmasked bad pixels in a single exposure will be removed when the images are shifted and co-added, greatly reducing the risk of failure. Such a strategy has already been adopted by some of JWST's near-infrared instruments. This change is currently being implemented, and will likely take effect in late 2026/early 2027.

\section{Time Series Observations using the Low-Resolution Slit Spectroscopy mode}\label{sec:extra}

An additional new capability is the implementation of TSOs using the LRS fixed-slit mode. The LRS mode is very popular for TSOs of transiting exoplanets in particular; until now, these have exclusively used the slitless LRS mode in the SLITLESSPRISM subarray. Performing lengthy TSOs in a very small aperture (the LRS slit measures $\sim$4.7 $\times$ 0.5'') risks introducing time-varying slit losses if the telescope pointing drifts over time. For this reason, TSOs were previously only permitted to use the slitless spectroscopy capability for MIRI's LRS mode (as well as Imaging). 

Many projects have, however, demonstrated that the pointing of JWST is exceptionally stable, with one study reporting a target remaining stable to $\leq$ 0.4\% of a pixel over a 24-hour observation\cite{2023arXiv230106350B}. This is easily stable enough to maintain placement inside the narrow slit. With the LRS slit mode offering improved sensitivity compared with the slitless mode (due to the focal plane mask suppressing the background around the source), a test observation was carried out to investigate the possibility of performing TSOs in the LRS slit. This analysis is the subject of a dedicated paper by Dyrek et al (2026), in these proceedings. 

TSOs using the LRS fixed-slit mode are further enabled by the new SUBSLIT subarray, which allows for faster frame times than the FULL array, which was the sole read-out option. Because of the placement and sizing of the SUBSLIT subarray, the slitless mode will still be optimal for the brightest targets.

\section{Summary}

JWST is currently entering its fifth year of operations; the observatory is delivering exceptional data quality and ground-breaking science. The MIRI instrument occupies a unique niche as the only instrument covering wavelengths longward of 5~\um. As we learn more about the instrument's performance and the needs of the astronomical community in getting the best science out of their data, we continue to explore new ways of delivering high-quality data. In this paper we describe several new features that are currently in different stages of implementation, that will expand the instrument's capabilities, improve data quality, and reduce the risk of failures. We anticipate making further such improvements to keep MIRI and JWST at the forefront of scientific discovery.

\bibliography{miri_tech} 
\bibliographystyle{spiebib} 

\end{document}